\documentclass[aps,showpacs,preprint]{revtex4}%
\usepackage{amsfonts}
\usepackage{amsmath}
\usepackage{amssymb}
\usepackage{graphicx}%
\begin{document}
\title{Orbital magnetization and its effect in antiferromagnets on the distorted fcc lattice}
\author{Zhigang Wang}
\affiliation{Institute of Applied Physics and Computational Mathematics, P.O. Box 8009,
Beijing 100088, P.R. China}
\author{Ping Zhang}
\thanks{Corresponding author. Electronic address: zhang\_ping@iapcm.ac.cn}
\affiliation{Institute of Applied Physics and Computational
Mathematics, P.O. Box 8009, Beijing 100088, P.R. China}
\author{Junren Shi}
\affiliation{Institute of Physics, P.O. Box 603, CAS, Beijing
100080, P.R. China}

\begin{abstract}
We study the intrinsic orbital magnetization (OM) in antiferromagnets on the
distorted face-centered-cubic lattice. The combined lattice distortion and
spin frustration induce nontrivial $k$-space Chern invariant, which turns to
result in profound effects on the OM properties. We derive a specific relation
between the OM and the Hall conductivity, according to which it is found that
the intrinsic OM vanishes when the electron chemical potential lies in the
Mott gap. The distinct behavior of the intrinsic OM in the metallic and
insulating regions is shown. The Berry phase effects on the thermoelectric
transport is also discussed.

\end{abstract}
\maketitle

\section{Introduction}

The orbital magnetism of Bloch electrons has been an outstanding problem in
solid state physics, and attracted renewed interest due to the recent
recognition \cite{Xiao1,Thon1,Xiao2} that the Berry phase effect plays a very
important role in the orbital magnetism. The issue was carried out in the
powerful semiclassical formalism \cite{Chang,Sund}, in which the Bloch
electron for $n$-th band is treated as a wave packet $|w_{n}(\mathbf{r}%
_{c},\mathbf{k}_{c})\rangle$ with its center ($\mathbf{r}_{c},\mathbf{k}_{c}$)
in the phase space. The orbital magnetic moment characterizes the rotation of
the wave packet around its centroid and is given by $\mathbf{m}_{n}%
(\mathbf{k}_{c})$=$\frac{(-e)}{2}\langle w_{n}|(\mathbf{\hat{r}}%
-\mathbf{r}_{c})\times\mathbf{\hat{v}}|w_{n}\rangle$, where $(-e)$ is the
charge of the electron and $\mathbf{\hat{v}}$ is the velocity operator. By
writing the wave packet in terms of the Bloch state, one obtains
($\mathbf{k}_{c}$ is abbreviated as $\mathbf{k}$)
\begin{equation}
\mathbf{m}_{n}(\mathbf{k})=-i(e/2\hbar)\langle\nabla_{\mathbf{k}%
}u_{n\mathbf{k}}|\times\lbrack\hat{H}_{\mathbf{k}}-\varepsilon_{n\mathbf{k}%
}^{(0)}]|\nabla_{\mathbf{k}}u_{n\mathbf{k}}\rangle, \label{e1}%
\end{equation}
where $|u_{n\mathbf{k}}\rangle$ is the periodic part of the Bloch state with
band energy $\varepsilon_{n\mathbf{k}}^{(0)}$, and $\hat{H}_{\mathbf{k}}$ is
the crystal Hamiltonian acting on $|u_{n\mathbf{k}}\rangle$. Equation
(\ref{e1}) can be alternatively derived by taking the differential of the
electron energy, which within first order in the perturbative magnetic field
$\mathbf{B}$ turns to be $\varepsilon_{n\mathbf{k}}$=$\varepsilon
_{n\mathbf{k}}^{(0)}-\mathbf{m}_{n}(\mathbf{k}){\small \cdot}\mathbf{B}$, with
respect to $\mathbf{B}$. It was further found \cite{Xiao1} that the presence
of a weak magnetic field $\mathbf{B}$ will result in a modification of the
density of states in the semiclassical phase space, $d^{3}\mathbf{k}%
\rightarrow d^{3}\mathbf{k}(1+e\mathbf{B}{\small \cdot}\mathbf{\Omega}%
_{n}/\hbar)$, where $\mathbf{\Omega}_{n}(\mathbf{k})=i\langle\nabla
_{\mathbf{k}}u_{n\mathbf{k}}|\times|\nabla_{\mathbf{k}}u_{n\mathbf{k}}\rangle$
is the Berry curvature in $k$-space. Due to this weak-field modification, a
quantum-state summation $\sum_{\mathbf{k}}\mathcal{O}(\mathbf{k})$ of some
physical quantity $\mathcal{O}(\mathbf{k})$ should be converted to an integral
according to $\int d^{3}\mathbf{k}(1+e\mathbf{B}{\small \cdot}\mathbf{\Omega
}_{n}/\hbar)\mathcal{O}(\mathbf{k})$. Additional thermodynamic average over
Bloch bands should be included at finite temperature. Therefore, the total
free energy for an equilibrium ensemble of electrons in the weak field may be
written as
\begin{equation}
F=-\frac{1}{\beta}\sum_{n}\int d^{3}\mathbf{k}\left(  1+\frac{e}{\hbar
}\mathbf{B}\cdot\mathbf{\Omega}_{n}(\mathbf{k})\right)  \ln[1+e^{\beta
(\mu-\varepsilon_{n\mathbf{k}})}]. \label{e2}%
\end{equation}
where $\mu$ is the electron chemical potential and $\beta=1/k_{B}T$. The
equilibrium orbital magnetization (OM) density is given by the field
derivative at fixed temperature and chemical potential, $\mathcal{\vec{M}%
}=-\left(  \partial F/\partial\mathbf{B}\right)  _{\mu,T}$, with the result%
\begin{align}
\mathcal{\vec{M}}  &  =\sum_{n}\int d^{3}\mathbf{km}_{n}(\mathbf{k}%
)f_{n}\label{e3}\\
&  +\frac{1}{\beta}\sum_{n}\int d^{3}\mathbf{k}\frac{e}{\hbar}\mathbf{\Omega
}_{n}(\mathbf{k})\ln\left[  1+e^{\beta(\mu-\varepsilon_{n\mathbf{k}})}\right]
\nonumber\\
&  \equiv\mathbf{M}_{c}+\mathbf{M}_{\mathbf{\Omega}},\nonumber
\end{align}
where $f_{n}$ is the local equilibrium Fermi function for $n$-th band. In
addition to the conventional term $\mathbf{M}_{c}$ in terms of the orbital
magnetic moment $\mathbf{m}_{n}(\mathbf{k})$, the extra term $\mathbf{M}%
_{\mathbf{\Omega}}$ in Eq. (\ref{e3}) is a Berry phase effect and exposes a
new topological ingredient to the orbital magnetism. Interestingly, it is this
Berry phase correction that eventually enters the thermal transport current
\cite{Xiao2}. At zero temperature and magnetic field the general expression
(\ref{e3}) is reduced to%
\begin{equation}
\mathcal{\vec{M}}=\sum_{n}\int^{\mu_{0}}d^{3}\mathbf{k}\left(  \mathbf{m}%
_{n}(\mathbf{k})+\frac{e}{\hbar}\mathbf{\Omega}_{n}(\mathbf{k})\left[  \mu
_{0}-\varepsilon_{n\mathbf{k}}\right]  \right)  , \label{e4}%
\end{equation}
where the upper limit means that the integral is over states with energies
below the zero-temperature chemical potential (Fermi energy) $\mu_{0}$.

The Berry phase effect on orbital magnetism was until now partially presented
by very few studies. Recent observation of the anomalous Nernst effect (ANE)
in CuCr$_{2}$Se$_{4-x}$Br$_{x}$ compound \cite{Lee} was attributed
\cite{Xiao2} to the manifestation of the Berry phase effect in the OM. Also
the orbital magnetism was recently studied by use of two-dimensional (2D)
Haldane model and ferromagnetic \textit{kagom\'{e}} lattice with spin
chirality \cite{Thon2,Wang2007}. These two models are rare examples to show
the zero-field quantum Hall effect (QHE) \cite{Haldane1988,Ohgushi}. From Ref.
\cite{Wang2007} one learns that the Berry phase effect causes the OM to
display different behavior in metallic and insulating regions. This difference
may be explained in parallel with Haldane's recent finding \cite{Haldane2004}
of the Berry phase effect in the intrinsic Hall conductivity [including QHE
and anomalous Hall effect (AHE)].

The objective of the present paper is dual. First we remark that the Berry
phase effect on the orbital magnetism has been included in the well-known
Kubo-Streda formula \cite{Streda1982}. Therefore, a full quantum-mechanical
linear response theory of the OM can be developed as a useful complement of
the semiclassical formalism, although the latter looks more elegant and
practical for calculation on clean samples. Then we examine the 3D problem by
studying the orbital magnetism in antiferromagnets on the distorted
face-centered-cubic (fcc) lattice. The results reveal that a general
\textquotedblleft\textit{topological orbital magnetism theory}%
\textquotedblright\ that takes into account Berry phase effect must now be developed.

The paper is organized as follows. In the next section, we address that the
intrinsic OM given in the semiclassical formalism is consistent with the
well-known quantum-mechanical Kubo-Streda formula in the clean-sample limit.
Section III describes the physical model that is used in this work. The
topological property and the consequent intrinsic Hall effect associated with
the model are also given in this section. In Sec. IV, we present a detailed
study of the properties of the OM and its effects on transport response in
antiferromagnets on the distorted fcc lattice. Finally, in Sec. V we present
our conclusions.

\section{Kubo-Streda formula of the orbital magnetization}

Due to the above mentioned modification of the density of states, the particle
number in the weak magnetic field (say, along $z$-axis) is given by
\begin{equation}
N(B,\mu)=\sum_{n}\int d^{3}\mathbf{k}\left(  1+\frac{e}{\hbar}B\Omega_{n}%
^{z}(\mathbf{k})\right)  f_{n}. \label{e5}%
\end{equation}
It is easy to see that a link between Eq. (\ref{e3}) and Eq. (\ref{e5}) is
$\left(  \frac{\partial\mathcal{M}_{z}}{\partial\mu}\right)  _{B}=\left(
\frac{\partial N}{\partial B}\right)  _{\mu}$, which is nothing but the usual
thermodynamic Maxwell relation and therefore should be free from the
weak-field limit used by the semiclassical approach. Thus the zero-field OM is
given by%
\begin{equation}
\mathcal{M}_{z}=\lim_{B\rightarrow0}\int^{\mu}\left(  \frac{\partial
N(B,\mu^{\prime})}{\partial B}\right)  _{\mu^{\prime}}d\mu^{\prime}.
\label{e6}%
\end{equation}
On the other side, the integrand in Eq. (\ref{e6}) can be written in terms of
Kubo-Streda \cite{Streda1982} formula for electrons as follows%
\begin{equation}
\sigma_{xy}|_{\mu}=\sigma_{xy}^{I}|_{\mu}-e\frac{\partial N(B,\mu)}{\partial
B}, \label{e7}%
\end{equation}
where $\sigma_{xy}|_{\mu}$ is the Hall conductivity and
\begin{equation}
\sigma_{xy}^{I}|_{\mu}=i\frac{e^{2}\hbar}{2}\int d\epsilon\frac{\partial
f\left(  \epsilon,\mu\right)  }{\partial\epsilon}\text{Tr}[\hat{v}_{x}%
G^{+}(\epsilon)\hat{v}_{y}\delta(\epsilon-\hat{H})-\hat{v}_{x}\delta
(\epsilon-\hat{H})\hat{v}_{y}G^{-}(\epsilon)]. \label{a1}%
\end{equation}
Here $G^{\pm}(\mu)=\lim_{\eta\rightarrow0^{+}}(\mu-\hat{H}\pm i\eta)^{-1}$ is
the operator Green function and $\hat{v}_{\alpha}$ is the velocity operator.
In Bloch-state representation the trace in Eq. (\ref{e6}) is equivalent to
$\sum_{n\mathbf{k}}\langle u_{n\mathbf{k}}|(\cdots)|u_{n\mathbf{k}}\rangle$
with the Hamiltonian transformed to $\hat{H}_{\mathbf{k}}=e^{i\mathbf{k}%
\cdot\mathbf{r}}\hat{H}e^{-i\mathbf{k}\cdot\mathbf{r}}$ and the velocity to
$\hat{v}_{\alpha}(\mathbf{k})$=$\frac{1}{\hbar}\frac{\partial\hat
{H}_{\mathbf{k}}}{\partial k_{\alpha}}$. Replacing $\frac{\partial f\left(
\epsilon,\mu\right)  }{\partial\epsilon}$ in Eq. (\ref{a1}) by $-\frac
{\partial f\left(  \epsilon,\mu\right)  }{\partial\mu}$ and using the
completeness relation of the Bloch states, $\sum_{nk}|u_{n\mathbf{k}}%
\rangle\langle u_{n\mathbf{k}}|$=$1$, one has
\begin{align}
\int^{\mu}\sigma_{xy}^{I}|_{\mu^{\prime}}d\mu^{\prime}  &  =-i\frac{e^{2}%
\hbar}{2}\lim_{\eta\rightarrow0^{+}}\int d\epsilon f(\epsilon,\mu)\times
\sum_{\substack{n,\mathbf{k}\\n^{\prime},\mathbf{k}^{\prime}}}\{\frac
{\delta(\epsilon-\varepsilon_{n\mathbf{k}})}{\epsilon-\varepsilon_{n^{\prime
}\mathbf{k}}+i\eta}\langle u_{n\mathbf{k}}|\hat{v}_{x}(\mathbf{k}%
)|u_{n^{\prime}\mathbf{k}^{\prime}}\rangle\langle u_{n^{\prime}\mathbf{k}%
^{\prime}}|\hat{v}_{y}(\mathbf{k})|u_{n\mathbf{k}}\rangle\nonumber\\
&  -\frac{\delta(\epsilon-\varepsilon_{n\mathbf{k}})}{\epsilon-\varepsilon
_{n^{\prime}\mathbf{k}^{\prime}}-i\eta}\langle u_{n\mathbf{k}}|\hat{v}%
_{y}(\mathbf{k})|u_{n^{\prime}\mathbf{k}^{\prime}}\rangle\langle u_{n^{\prime
}\mathbf{k}^{\prime}}|\hat{v}_{x}(\mathbf{k})|u_{n\mathbf{k}}\rangle
\}\nonumber\\
&  =\frac{e^{2}}{\hbar}\sum_{\substack{n,\mathbf{k}\\n^{\prime},\mathbf{k}%
^{\prime}}}f(\varepsilon_{n\mathbf{k}},\mu)\frac{\operatorname{Im}\{\langle
u_{n\mathbf{k}}|\frac{\partial\hat{H}_{\mathbf{k}}}{\partial k_{x}%
}|u_{n^{\prime}\mathbf{k}^{\prime}}\rangle\langle u_{n^{\prime}\mathbf{k}%
^{\prime}}|\frac{\partial\hat{H}_{\mathbf{k}}}{\partial k_{y}}|u_{n\mathbf{k}%
}\rangle\}}{\varepsilon_{n\mathbf{k}}-\varepsilon_{n^{\prime}\mathbf{k}%
^{\prime}}}. \label{a2}%
\end{align}
By use of the identity
\begin{equation}
\frac{\partial\hat{H}_{\mathbf{k}}}{\partial k_{\alpha}}|u_{n\mathbf{k}%
}\rangle=\frac{\partial\varepsilon_{n\mathbf{k}}}{\partial k_{\alpha}%
}|u_{n\mathbf{k}}\rangle+(\varepsilon_{n\mathbf{k}}-\hat{H}_{\mathbf{k}%
})|\frac{\partial u_{n\mathbf{k}}}{\partial k_{\alpha}}\rangle\label{a3}%
\end{equation}
and after a transformation of $\mathbf{k}$-sum to an integral, Eq. (\ref{a2})
is ready to be simplified as
\begin{equation}
\int^{\mu}\sigma_{xy}^{I}|_{\mu^{\prime}}d\mu^{\prime}=\frac{e^{2}}{\hbar}%
\sum_{n}\int d^{3}\mathbf{k}f_{n}\operatorname{Im}\left\{  \langle
\frac{\partial u_{n\mathbf{k}}}{\partial k_{x}}|\hat{H}_{\mathbf{k}%
}-\varepsilon_{n\mathbf{k}}|\frac{\partial u_{n\mathbf{k}}}{\partial k_{y}%
}\rangle\right\}  . \label{a4}%
\end{equation}
A comparison of Eq. (\ref{a4}) with Eq. (\ref{e1}) immediately gives the
following relation%
\begin{align}
\int^{\mu}\sigma_{xy}^{I}|_{\mu^{\prime}}d\mu^{\prime}  &  =e\sum_{n}\int
d^{3}\mathbf{k}m_{n}^{z}(\mathbf{k})f_{n}\label{a5}\\
&  =eM_{c}^{(z)},\nonumber
\end{align}
where the second line is obtained by using the definition of $\mathbf{M}_{c}$
in Eq. (\ref{e3}). Thus one finds that the semiclassical expression for
$\mathbf{M}_{c}$ is equivalent to the quantum-mechanical expression for
$(1/e)\int^{\mu}\sigma_{xy}^{I}|_{\mu^{\prime}}d\mu^{\prime}$. Note that
although $\sigma_{xy}^{I}$ is a Fermi-surface term, the quantity $\int^{\mu
}\sigma_{xy}^{I}|_{\mu^{\prime}}d\mu^{\prime}$ is a Fermi-sea term and all the
Bloch states below $\mu$ should be accounted when calculating the OM. On the
other side, in the clean limit, the Kubo formula for the Hall conductivity
$\sigma_{xy}|_{\mu}$ can be written in terms of the Berry curvatures
$\Omega_{n}^{z}$ \cite{Thouless1982},
\begin{equation}
\sigma_{xy}|_{\mu}=-\frac{e^{2}}{\hbar}\sum_{n}\int d^{3}\mathbf{k}f_{n}%
\Omega_{n}^{z}. \label{a6}%
\end{equation}
From Eq. (\ref{a6}) one has%
\begin{equation}
\int^{\mu}\sigma_{xy}|_{\mu^{\prime}}d\mu^{\prime}=-\frac{1}{\beta}\frac
{e^{2}}{\hbar}\sum_{n}\int d^{3}\mathbf{k\Omega}_{n}(\mathbf{k})\ln\left[
1+e^{\beta(\mu-\varepsilon_{n\mathbf{k}})}\right]  , \label{a7}%
\end{equation}
which is exactly the semiclassical expression for the Berry phase term
$eM_{\Omega}^{(z)}$ in Eq. (\ref{e3}). Thus the semiclassical OM in Eq.
(\ref{e3}) can actually be written as a Kubo-Streda formula:%
\begin{equation}
\mathcal{M}_{z}=\frac{1}{e}\int^{\mu}\sigma_{xy}^{I}|_{\mu^{\prime}}%
d\mu^{\prime}-\frac{1}{e}\int^{\mu}\sigma_{xy}|_{\mu^{\prime}}d\mu^{\prime}.
\label{a8}%
\end{equation}
The other two components $\mathcal{M}^{x}$ and $\mathcal{M}^{y}$ are given in
a similar manner. This equivalence between the semiclassical and
quantum-mechanical description for the OM is only valid in the intrinsic
region and will break down when the impurity scattering effect is included.
Thus while the semiclassical formula of the OM is more suitably employed to
study the intrinsic property of the OM, the Kubo-Streda formula must be used
when one takes into account the impurity scattering. Another aspect is that
the Kubo-Streda formula is valid in arbitrary strength of the external
magnetic field, while the semiclassical formula only works in the weak field
limit. In some special cases, for example, when one wants to know the edge
state effect on the OM in a finite-size sample in a strong magnetic field $B$
\cite{Wak1999,Streda1994}, a more apparent approach can be employed by
directly calculating the total free energy and then a finite-field OM is
obtained by a $B$-derivative of the free energy.

\section{Specific model and 2D Chern number}

Now we focus our attention to the properties of the OM in a specific
spin-frustrated system. As in Ref. \cite{Shindou2001}, the model we used
describes the chiral spin state in the ordered antiferromagnet (AF) on the
three-dimensional fcc lattice. The AF on the fcc lattice is a typical
frustrated system, and nontrivial triple-$Q$ spin structure with finite spin
chirality has been revealed by band structure calculation \cite{Sakuma2000}
and observed in experiments \cite{Wilson,Endoh}. The anomalous behaviors in
the fcc AF were also observed. For example, there occurs mysterious weak
ferromagnetism in NiS$_{2}$ below the second AF transition temperature
\cite{Thio1995}. The Hall conductivity in this material is also large and
strongly temperature dependent \cite{Thio1994}. In Co(S$_{x}$Se$_{1-x}$)$_{2}%
$, the AHE is enhanced in the intermediate $x$ region, where the nontrivial
magnetism is realized \cite{Ada1981}.%

\begin{figure}[tbp]
\begin{center}
\includegraphics[width=0.6\linewidth]{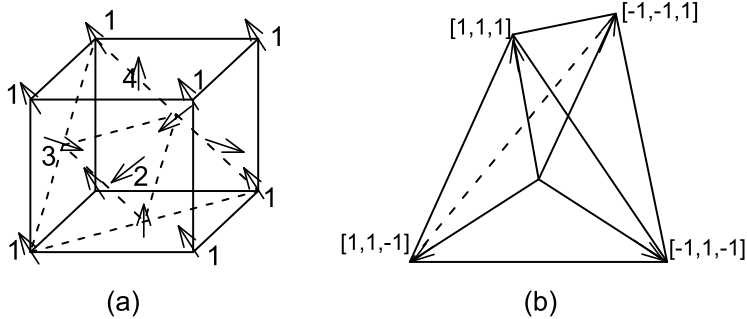}
\end{center}
\caption{(a) Triple-$Q$ spin structure on fcc lattice. (b) Relation of the
$4$-spin moments $\vec{S}_{a}$ ($a=1,2,3,4$).}
\label{fig1}
\end{figure}%
The triple-$Q$ spin structure on facc lattice is shown in Fig. 1. Here the
lattice points are divided into four sublattices with different local spins
$\vec{S}_{a}$ ($a=1,2,3,4$) on them. The AF nature requires $\sum_{a}\vec
{S}_{a}$=$0$. The minimization of the 2-spin exchange interaction energy
cannot uniquely determine the sublattice spin orientation. The inclusion of
higher order (4-spin exchange) interaction $H_{4}=J_{4}\sum_{a\neq b}(\vec
{S}_{a}{\small \cdot}\vec{S}_{b})^{2}$ with positive $J_{4}$ gives the
ground-state spin configuration \cite{Yoshida1980,Yoshimori1981} as $\vec
{S}_{1}$=$(\frac{1}{\sqrt{3}},\frac{1}{\sqrt{3}},\frac{1}{\sqrt{3}})$,
$\vec{S}_{2}$=$(\frac{1}{\sqrt{3}},-\frac{1}{\sqrt{3}},-\frac{1}{\sqrt{3}})$,
$\vec{S}_{3}$=$(-\frac{1}{\sqrt{3}},\frac{1}{\sqrt{3}},-\frac{1}{\sqrt{3}})$,
and $\vec{S}_{4}$=$(-\frac{1}{\sqrt{3}},-\frac{1}{\sqrt{3}},\frac{1}{\sqrt{3}%
})$, where each direction corresponds to the four corners from the center of a
tetrahedron [Fig. 1(b)]. The effective Hamiltonian for the hopping electrons
strongly coupled to the mean-field effective magnetic field caused by these
local spins is given by $H=\sum_{NN}t_{ij}^{eff}\psi_{i}^{\dag}\psi_{j}$ with
$t_{ij}^{eff}=t\langle\chi_{i}|\chi_{j}\rangle=te^{ia_{ij}}\cos\frac
{\vartheta_{ij}}{2}$. Here the spin wave function $|\chi_{i}\rangle$ is
explicitly given by $|\chi_{i}\rangle=\left[  \cos\frac{\vartheta_{i}}%
{2},\text{ }e^{i\phi_{i}}\sin\frac{\vartheta_{i}}{2}\right]  ^{\text{T}}$,
where the polar coordinates are pinned by the local spins, i.e., $\langle
\chi_{i}|\vec{S}_{i}|\chi_{i}\rangle=\frac{1}{2}\left(  \sin\vartheta_{i}%
\cos\phi_{i},\text{ }\sin\vartheta_{i}\sin\phi_{i},\text{ }\cos\vartheta
_{i}\right)  $. $\vartheta_{ij}$ is the angle between the two spins $\vec
{S}_{i}$ and $\vec{S}_{j}$. The phase factor $a_{ij}$ can be regarded as the
gauge vector potential $a_{\mu}(\mathbf{r})$, and the corresponding gauge flux
is related to scalar spin chirality $\chi_{ijk}$=$\vec{S}_{i}{\small \cdot
}(\vec{S}_{j}{\small \times}\vec{S}_{k})$ \cite{Laughlin}. In periodic crystal
lattices, the non-vanishing of the gauge flux relies on the multiband
structure with each band being characterized by a Chern number. The Chern
number appears as a result of the spin-orbit interaction and/or spin chirality
in ferromagnets. In ferromagnets the time-reversal broken symmetry is
manifest, while in AF the time-reversal operation combined with the
translation operation often constitutes the unbroken symmetry. In the latter
case, the nonzero Hall conductivity $\sigma_{xy}$ is forbidden. However, when
there are more than two sublattices and the spin structure is noncollinear,
this combined symmetry would be absent and finite $\sigma_{xy}$ is not
forbidden \cite{Shindou2001}.

The net spin chirality for the ideal fcc AF lattice in Fig. 1 is zero, because
the spin chiralities are the vector quantities and the sum of these four
vectors is zero. However, when the lattice is distorted along the [1,1,1]
direction, then the non-zero net spin chirality occurs. Following Ref.
\cite{Shindou2001}, we express the distortion along the [1,1,1] direction by
putting the transfer integral within the (1,1,1) plane as $t_{\text{intra}}%
$=1, while that between the planes as $t_{\text{inter}}$=1$-d$. As the unit
cell is cubic shown in Fig. 1, the first Brillouin zone (BZ) is cubic:
[$-\frac{\pi}{a},\frac{\pi}{a}$]$^{3}$. From now on, we set $a$=$1$. Then the
Hamiltonian matrix $H_{\mathbf{k}}$ for each $\mathbf{k}$ is given by%
\begin{equation}
H_{\mathbf{k}}=\left(
\begin{array}
[c]{cccc}%
0 & e^{-i\frac{\pi}{6}}f_{2} & e^{i\frac{\pi}{6}}f_{1} & f_{3}\\
e^{i\frac{\pi}{6}}f_{2} & 0 & e^{-i\frac{\pi}{6}}f_{3} & e^{i\frac{2\pi}{3}%
}f_{1}\\
e^{-i\frac{\pi}{6}}f_{1} & e^{i\frac{\pi}{6}}f_{3} & 0 & e^{-i\frac{2\pi}{3}%
}f_{2}\\
f_{3} & e^{-i\frac{2\pi}{3}}f_{1} & e^{i\frac{2\pi}{3}}f_{2} & 0
\end{array}
\right)  , \label{e10}%
\end{equation}
where $f_{1}$=$2(1-d)\cos(\frac{k_{z}}{2}$+$\frac{k_{x}}{2})$+$2\cos
(-\frac{k_{z}}{2}$+$\frac{k_{x}}{2})$, $f_{2}$=$2(1-d)\cos(\frac{k_{x}}{2}%
$+$\frac{k_{y}}{2})$+$2\cos(-\frac{k_{x}}{2}$+$\frac{k_{y}}{2})$, $f_{3}%
$=$2(1-d)\cos(\frac{k_{y}}{2}$+$\frac{k_{z}}{2})$+$2\cos(-\frac{k_{y}}{2}%
$+$\frac{k_{z}}{2})$. In this Hamiltonian, the two lower bands are fully
degenerate, $\varepsilon_{1\mathbf{k}}$=$\varepsilon_{2\mathbf{k}}%
(\mathbf{k})$=$-\sqrt{f_{1}^{2}+f_{2}^{2}+f_{3}^{2}}$, while the two upper
bands are also degenerate with $\varepsilon_{3,4\mathbf{k}}$=$\sqrt{f_{1}%
^{2}+f_{2}^{2}+f_{3}^{2}}$. The band structure along high-symmetry lines in
the first BZ is shown in Fig. 2.%
\begin{figure}[tbp]
\begin{center}
\includegraphics[width=1.0\linewidth]{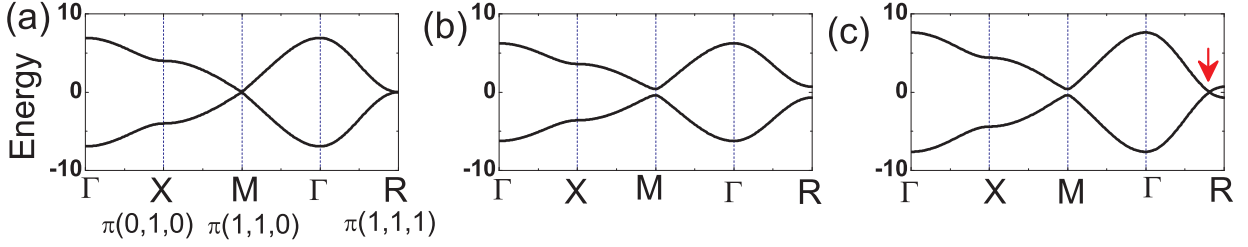}
\end{center}
\caption{Energy spectrum along high-symmetry lines in the first BZ for (a)
$d$=$0$, (b) $d$=$0.2$, and (c) $d$=$-0.2$.}
\label{fig2}
\end{figure}
At $d$=$0$ [Fig. 2(a)], the upper and lower dispersions touch along the edge
of the BZ and comprises an assembly of the massless Dirac fermions (Weyl
fermions) in $(2+1)$D. For $d$%
$>$%
$0$ (elongation along [1,1,1] direction), all the Weyl fermions along the edge
open a gap and turn into the massive Dirac fermions [Fig. 2(b)]. Therefore the
gap fully opens in the density of states centered at zero energy. For $d$%
$<$%
$0$ (suppression along [1,1,1] direction), all the $(2+1)$D Weyl fermions
along the edges open the gap as in the case of $d$%
$>$%
$0$. However, there occurs two additional $(3+1)$D Weyl fermions [Fig. 2(c)]
at $(k_{x},k_{y},k_{z})=\pm D(1,1,1)$, where $D\equiv\arccos\left(  \frac
{1}{d-1}\right)  $ and $\pm$ correspond to the right- and left-handed
chirality \cite{Nie1983}. Thus unlike $d$%
$>$%
$0$, the gap does not fully open in the case of $d$%
$<$%
$0$ due to the presence of a new single contact between the upper and lower
bands in the BZ. The normalized eigenvectors are given by%
\begin{align}
|u_{1\mathbf{k}}\rangle &  =\frac{1}{\sqrt{2}}\left(
\begin{array}
[c]{cccc}%
-\frac{f_{1}\varepsilon_{1\mathbf{k}}+if_{2}f_{3}}{\varepsilon_{1\mathbf{k}%
}\sqrt{f_{1}^{2}+f_{2}^{2}}}, & e^{-i\frac{\pi}{3}}\frac{f_{2}\varepsilon
_{1\mathbf{k}}-if_{1}f_{3}}{\varepsilon_{1\mathbf{k}}\sqrt{f_{1}^{2}+f_{2}%
^{2}}}, & 0, & -\frac{\sqrt{f_{1}^{2}+f_{2}^{2}}}{\varepsilon_{1\mathbf{k}}}%
\end{array}
\right)  ^{\text{T}},\label{e11}\\
|u_{2\mathbf{k}}\rangle &  =\frac{1}{\sqrt{2}}\left(
\begin{array}
[c]{cccc}%
e^{i\frac{\pi}{6}}\frac{f_{2}}{\varepsilon_{2\mathbf{k}}}, & e^{-i\frac{\pi
}{6}}\frac{f_{1}}{\varepsilon_{2\mathbf{k}}}, & 1, & ie^{i\frac{\pi}{6}}%
\frac{f_{3}}{\varepsilon_{2\mathbf{k}}}%
\end{array}
\right)  ^{\text{T}},\nonumber\\
|u_{3\mathbf{k}}\rangle &  =\frac{1}{\sqrt{2}}\left(
\begin{array}
[c]{cccc}%
\frac{f_{1}\varepsilon_{3\mathbf{k}}+if_{2}f_{3}}{\varepsilon_{3\mathbf{k}%
}\sqrt{f_{1}^{2}+f_{2}^{2}}}, & -e^{-i\frac{\pi}{3}}\frac{f_{2}\varepsilon
_{3\mathbf{k}}-if_{3}f_{1}}{\varepsilon_{3\mathbf{k}}\sqrt{f_{1}^{2}+f_{2}%
^{2}}}, & 0, & \frac{\sqrt{f_{1}^{2}+f_{2}^{2}}}{\varepsilon_{3\mathbf{k}}}%
\end{array}
\right)  ^{\text{T}},\nonumber\\
|u_{4\mathbf{k}}\rangle &  =\frac{1}{\sqrt{2}}\left(
\begin{array}
[c]{cccc}%
e^{i\frac{\pi}{6}}\frac{f_{2}}{\varepsilon_{4\mathbf{k}}}, & e^{-i\frac{\pi
}{6}}\frac{f_{1}}{\varepsilon_{4\mathbf{k}}}, & 1, & ie^{i\frac{\pi}{6}}%
\frac{f_{3}}{\varepsilon_{4\mathbf{k}}}%
\end{array}
\right)  ^{\text{T}}.\nonumber
\end{align}
The Berry curvatures for these four Bloch bands are derived to have the form
\begin{equation}
\Omega_{n}^{\alpha}(\mathbf{k})=\frac{F_{\beta\gamma}}{2\varepsilon
_{n\mathbf{k}}^{3}} \label{e12}%
\end{equation}
with
\begin{equation}
F_{\beta\gamma}=f_{1}\frac{\partial f_{2}}{\partial k_{\beta}}\frac{\partial
f_{3}}{\partial k_{\gamma}}+f_{3}\frac{\partial f_{1}}{\partial k_{\beta}%
}\frac{\partial f_{2}}{\partial k_{\gamma}}-f_{2}\frac{\partial f_{1}%
}{\partial k_{\beta}}\frac{\partial f_{3}}{\partial k_{\gamma}}, \label{e13}%
\end{equation}
where ($\alpha,\beta,\gamma$) represent a cyclic permutation of ($x,y,z$).

Let us see the Hall conductivity of this system \cite{Shindou2001} with
$d\neq0$. In the integer filling case, the zero-temperature Hall conductivity
is a sum of Chern invariant \cite{Koh1992} over occupied Bloch bands,
\begin{equation}
\sigma_{xy}=(e^{2}/h)\sum_{n}^{\text{occu}}\int_{[-\pi:\pi]}\frac{dk_{z}}%
{2\pi}C_{n}(k_{z}), \label{e14}%
\end{equation}
where the 2D Chern number \cite{Thouless} $C_{n}(k_{z})$ is given by
\begin{align}
C_{n}(k_{z})  &  =-\frac{1}{2\pi}\int_{[-\pi:\pi]^{2}}dk_{x}dk_{y}\Omega
_{n}^{z}(\mathbf{k})\label{e15}\\
&  =-\frac{1}{2\pi}\int_{[-\pi:\pi]^{2}}dk_{x}dk_{y}\text{ }\hat{z}%
\cdot\left(  \nabla_{\mathbf{k}}\times\mathbf{A}_{n}(\mathbf{k})\right)
.\nonumber
\end{align}
Here $\mathbf{A}_{n}(\mathbf{k})$=$i\langle u_{n\mathbf{k}}|\nabla
_{\mathbf{k}}u_{n\mathbf{k}}\rangle$ is the Berry phase connection (vector
potential) for $n$-th band. To proceed one may first transform the integral of
$\nabla_{\mathbf{k}}\times\mathbf{A}_{n}$ over the first BZ to the line
integral of $\mathbf{A}_{n}$ along the BZ boundary by use of Stokes' theorem,
and then apply the complex contour integration technique and residue theorem
to sinusoidal functions. After a straightforward derivation, one obtains the
non-zero Chern number, $C_{1}(k_{z})=-$sgn$(g(k_{z}))$, $C_{3}=$%
sgn$(g(k_{z}))$, where%
\begin{equation}
g(k_{z})=2+2(1-d)\cos(k_{z}+2k_{P}) \label{e16}%
\end{equation}
and $k_{P}=\arctan\left[  \frac{(d-1)\cos k_{z}-1}{(d-1)\sin k_{z}}\right]  $.
It is easy to verify that for $d$%
$>$%
$0$, the value of $g(k_{z})$ is always positive, independent of $k_{z}$. For
$d$%
$<$%
$0$, the sign of $g(k_{z})$ depends on $k_{z}$ in such a way that $g(k_{z})$%
$<$%
$0$ for $k_{z}\in(-D,D)$ and $g(k_{z})$%
$>$%
$0$ for $k_{z}\in\lbrack-\pi,-D)\cup(D,\pi]$. Note that the present choice of
the other two Bloch states $|u_{2\mathbf{k}}\rangle$ and $|u_{4\mathbf{k}%
}\rangle$ makes them to have no contribution to the Chern number.

However, the above purely mathematical calculation of 2D Chern number is not
favored by theoretical physicists, who would like to resort to the physical
connotation that the vector potential $\mathbf{A}_{n}$ and gauge flux
$\mathbf{\Omega}_{n}$ are endowed with. Correspondingly, here we present this
gauge-field analysis of the lower band $\varepsilon_{1\mathbf{k}}$ as an
example. The value of 2D Chern number $C_{1}(k_{z})$, which is confined to the
($k_{x},k_{y}$) subspace at fixed $k_{z}$, is invariant under gauge
transformation $|u_{1\mathbf{k}}^{\prime}\rangle$=$e^{i\varphi_{1}%
(\mathbf{k})}|u_{1\mathbf{k}}\rangle$, $\mathbf{A}_{1}^{\prime}(\mathbf{k}%
)=\mathbf{A}_{1}(\mathbf{k})-\nabla_{\mathbf{k}}\varphi_{1}(\mathbf{k})$,
where $\varphi_{1}(\mathbf{k})$ is an arbitrary smooth function of
$\mathbf{k}$. If the gauge choice for $|u_{1\mathbf{k}}\rangle$ is
well-defined everywhere in the whole ($k_{x},k_{y}$) subspace in the first BZ,
then its Chern number $C_{1}(k_{z})$ will obviously be zero. However, at point
$\mathbf{k}_{0}=(k_{z}+2k_{P},k_{z},k_{z})$, one can find that the wave
function $|u_{1\mathbf{k}}\rangle$ in Eq. (\ref{e11}) is ill-defined since
both its denominator and numerator are zero at this point. This means that the
used gauge cannot apply to the whole BZ and one needs to render a gauge
transformation to avoid the singularity at $\mathbf{k}_{0}$. For this one
transforms the wave function to
\begin{equation}
|u_{1\mathbf{k}}^{\prime}\rangle=\frac{1}{\sqrt{2}}\left(
\begin{array}
[c]{cccc}%
-\frac{f_{2}\varepsilon_{1\mathbf{k}}+if_{1}f_{3}}{\varepsilon_{1\mathbf{k}%
}\sqrt{f_{2}^{2}+f_{3}^{2}}}, & e^{-i\frac{\pi}{3}}\frac{f_{3}\varepsilon
_{1\mathbf{k}}-if_{1}f_{2}}{\varepsilon_{1\mathbf{k}}\sqrt{f_{2}^{2}+f_{3}%
^{2}}}, & 0, & -\frac{\sqrt{f_{2}^{2}+f_{3}^{2}}}{\varepsilon_{1\mathbf{k}}}%
\end{array}
\right)  ^{\text{T}}. \label{e17}%
\end{equation}
The new eigenvector $|u_{1\mathbf{k}}^{\prime}\rangle$ recovers the
well-defined behavior at $\mathbf{k}_{0}$; the new singularity brought about
is at $\mathbf{k}_{0}^{\prime}=(k_{z},k_{z}+2k_{P},k_{z})$. Thus according to
the two different gauge choices, the BZ cross section at fixed $k_{z}$ is now
divided into two regions V and V$^{\prime}$ as shown in Fig. 3 ($k_{z}$=$0$)%
\begin{figure}[tbp]
\begin{center}
\includegraphics[width=0.6\linewidth]{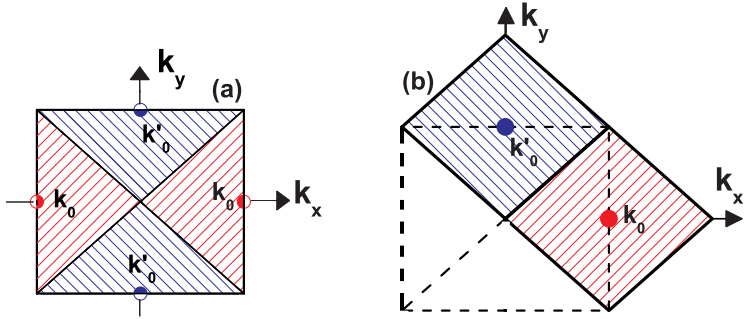}
\end{center}
\caption{(Color online). Division of cross section ($k_{z}%
$=$0$) of the first BZ
into two regions V (red area) and V$^{\prime}$ (blue area). Note that the
projection of the singularities $\mathbf{k}_{0}$ and $\mathbf{k}_{0}^{\prime}$
onto $(k_{x},k_{y})$ subspace vary with $k_{z}$.}
\label{fig3}
\end{figure}%
. The wave functions $|u_{1\mathbf{k}}\rangle$ are used onto the region V,
while $|u_{1\mathbf{k}}^{\prime}\rangle$ apply to V$^{\prime}$. Note that
there remains some freedom in the division of the BZ. Because $|u_{1\mathbf{k}%
}\rangle$ and $|u_{1\mathbf{k}}^{\prime}\rangle$ are ill-defined only at
$\mathbf{k}_{0}$ and $\mathbf{k}_{0}^{\prime}$, respectively, we are free to
deform this division as long as $\mathbf{k}_{0}$($\mathbf{k}_{0}^{\prime}%
$)$\notin$V(V$^{\prime}$). This corresponds to the gauge degree of freedom
\cite{Koh1985,Muk2003}. At \textbf{k}$\mathbf{\in}$V$\cap$V$^{\prime}$, the
two choices of wave functions are different by a phase factor $|u_{1\mathbf{k}%
}^{\prime}\rangle$=$e^{i\varphi_{1}(\mathbf{k})}|u_{1\mathbf{k}}\rangle$,
i.e., $\mathbf{A}_{1}^{\prime}(\mathbf{k})$=$\mathbf{A}_{1}(\mathbf{k}%
)-\nabla_{\mathbf{k}}\varphi_{1}(\mathbf{k})$, where
\begin{equation}
e^{i\varphi_{1}(\mathbf{k})}=\sqrt{\frac{f_{2}^{2}+f_{1}^{2}}{f_{3}^{2}%
+f_{2}^{2}}}\frac{f_{2}\varepsilon_{1\mathbf{k}}+if_{3}f_{1}}{f_{1}%
\varepsilon_{1\mathbf{k}}+if_{2}f_{3}}. \label{e18}%
\end{equation}
Thus one obtains the value of nonzero 2D Chern number for lower band
$\varepsilon_{1}(\mathbf{k})$ as follows
\begin{align}
C_{1}(k_{z})  &  =-\frac{1}{2\pi}\oint_{\partial\text{V}}\left[
\mathbf{A}_{1}(\mathbf{k})-\mathbf{A}_{1}^{\prime}(\mathbf{k})\right]
{\small \cdot}d\mathbf{k}\label{e19}\\
&  =-\frac{1}{2\pi}\oint_{\partial\text{V}}d\varphi_{1}(\mathbf{k}%
)=-\text{sgn}(g(k_{z})),\nonumber
\end{align}
which is consistent with the explicit calculation based on the complex-contour
integration technique.

Consider the $\mu$=$0$ case, i.e., the two lower degenerate bands are fully
filled while the two upper degenerate bands are empty. Then a further $k_{z}%
$-integral of $C_{1}(k_{z})$ gives the Hall conductivity $\sigma_{xy}%
=-\frac{e^{2}}{h}$ for $d$%
$>$%
$0$ and $\sigma_{xy}=\frac{e^{2}}{h}\left(  \frac{2D}{\pi}-1\right)  $ for $d$%
$<$%
$0$. The asymmetry of $\sigma_{xy}$ between $d$%
$>$%
$0$ and $d$%
$<$%
$0$ will be explained below together with the behavior of the OM. When the
local spins $\{\vec{S}_{i}\}$ are inverted (which means that the spin
chirality is also inverted), then the Hall conductivity changes its sign.

\section{Orbital magnetization and its effects}

Now we turn to study the OM and its various effects. Without loss of
generality, the present attention is only on the $z$-component of the OM which
is connected with Hall conductivity $\sigma_{xy}$. First, after a
straightforward derivation, one obtains the $k$-space orbital magnetic moment
as follows
\begin{equation}
m_{n}^{z}(\mathbf{k})=\frac{F_{xy}}{2\varepsilon_{n}^{2}(\mathbf{k})}.
\label{e20}%
\end{equation}
See Eq. (\ref{e13}) for $F_{xy}$. One can see that the orbital magnetic moment
is identical for upper and lower bands, while the Berry curvatures [Eq.
(\ref{e12})] for upper and lower bands differ by a sign. A comparison between
Eq. (\ref{e20}) and Eq. (\ref{e12}) gives an interesting relation for the
present model%
\begin{equation}
m_{n}^{z}(\mathbf{k})=\Omega_{n}^{z}(\mathbf{k})\varepsilon_{n\mathbf{k}}.
\label{e21}%
\end{equation}
Given the expressions for $m_{n}^{z}(\mathbf{k})$ and $\Omega_{n}%
^{z}(\mathbf{k})$, the OM $\mathcal{M}^{z}$ can now be systematically studied.
At zero temperature, in particular, by substituting Eq. (\ref{e21}) into Eq.
(\ref{e4}) one finds an important relation%
\begin{align}
\mathcal{M}_{z}  &  =\frac{e}{\hbar}\mu_{0}\sum_{n}\int^{\mu_{0}}%
d^{3}\mathbf{k}\Omega_{n}^{z}(\mathbf{k})\label{e22}\\
&  =-\frac{\mu_{0}}{e}\sigma_{xy},\nonumber
\end{align}
which indicates that the OM is proportional to the Hall conductivity with
coefficient ($-\mu_{0}/e$). Equation (\ref{e22}) also holds at low
temperature. Although this remarkable relation between the OM and the Hall
conductivity is specific to the present Hamiltonian model, it definitely tells
one that the topological ingredient in the OM may be faithfully mapped out
through the Hall conductivity. If the band is partially filled, then after
integrating by parts one finds that Eq. (\ref{e22}) can be written as a pure
Fermi-surface integral. In the integer filling case, on the other hand, the OM
and Hall conductivity display a Fermi-sea feature. Figure 4 shows
$\mathcal{M}_{z}$ and $\sigma_{xy}$ as a function of the distortion $d$.%
\begin{figure}[tbp]
\begin{center}
\includegraphics[width=0.6\linewidth]{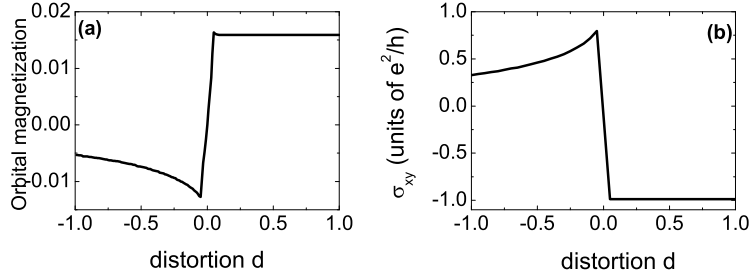}
\end{center}
\caption{(a) Zero-temperature orbital magnetization $\mathcal{M}_{z}$ and (b)
Hall conductivity $\sigma_{xy}$ as a function of distortion $d$. The Fermi
energy in (a) is chosen to be $\mu_{0}$=$0.1$, while in (b) the Fermi energy
is chosen to be $\mu_{0}$=$0.0$ which corresponds to case that only the lower
bands are fully occupied.}
\label{fig4}
\end{figure}
The $d$%
$>$%
$0$ and $d$%
$<$%
$0$ cases are asymmetric by the observation that $\sigma_{xy}$ is quantized
for $d$%
$>$%
$0$ while non-quantized for $d$%
$<$%
$0$. To understand this asymmetry one may treat the distortion $d$ as a
control parameter of the 3D band structure and start from $d$=$0$, at which
the upper and lower bands are degenerate along the BZ edge, i.e., ($k_{x}%
$=$\pm\pi,k_{y}$=$\pm\pi,k_{z}$), ($k_{x}$=$,k_{y}$=$\pm\pi,k_{z}$=$\pm\pi$),
($k_{x}$=$\pm\pi,k_{y},k_{z}$=$\pm\pi$). When $d$ is varied from $d$=$0$ to
$d$%
$>$%
$0$, then these initial degenerate points \textit{completely} split into two
groups of \textquotedblleft Dirac point\textquotedblright\ singularities,
which play the role of positive and negative monopole sources, respectively
[see, as an example, $\mathbf{k}_{0}$ and $\mathbf{k}_{0}^{\prime}$ points in
Fig. 2]. The upper and lower bands are now tightly coupled by a series of
\textquotedblleft Berry flux loops\textquotedblright. Along each loop Berry
curvature flux $2\pi$ passes from the lower bands to the upper bands through
one Dirac point, then returns through the other corresponding one. The
positive (negative) monopoles of the lower bands and the negative (positive)
monopoles of the upper bands may recombine by a relative displacement of a
primitive reciprocal lattice vector $\mathbf{G}$. In the present cubic lattice
the $\mathbf{G}$ is along one selective $\hat{k}_{\alpha}$-axis ($\alpha
=x,y,z$) with amplitude $2\pi$ (the lattice constant $a$ has been scaled to be
unity). This means that during the \textquotedblleft Dirac
point\textquotedblright\ splitting process, the individual Chern invariant
(the 3D generalization of 2D Chern number) for the lower and upper bands
changes by $\mp\mathbf{G}$, respectively, while their sum is conserved. As a
result, the Hall conductivity $\sigma_{xy}$ for lower/upper bands is quantized
to be a product of $\frac{e^{2}}{h}$ and the following Chern invariant%
\begin{equation}
\frac{1}{2\pi}\int_{[-\pi:\pi]^{3}}d^{3}\mathbf{k}\Omega_{n}^{z}%
(\mathbf{k})=-\frac{1}{2\pi}C_{n}G_{z}. \label{e23}%
\end{equation}
When $d$ is varied from $d$=$0$ to $d$%
$<$%
$0$, equation (\ref{e23}) breaks down, since the 2D Chern number $C_{n}$ is
now $k$-dependent by the presence of the additional two (3+1)D Weyl fermions
at $(k_{x},k_{y},k_{z})=\pm D(1,1,1)$. In this case, although the gap is
opened at Dirac points along the BZ edge, the emergence of new Weyl fermions
contributes non-zero density of states in the gap. It is then straightforward
to repeat the integration in Eq. (\ref{e23}) by parts to expose the
non-quantized part of the 3D intrinsic Hall conductivity as a Fermi surface property.

In the half filled case, i.e., when the chemical potential is in the Mott gap,
the Hall conductivity vanishes due to the cancellation of the Chern invariants
of the upper and lower bands. As a consequence, the OM in Eq. (\ref{e23}) also
vanishes. This result is prominently different from that in Ref.
\cite{Shindou2001}, in which a finite OM with amplitude smoothly varied with
distortion $d$ was given via a tight-binding calculation. Ref.
\cite{Shindou2001} employed the Kubo-Streda formula in calculating the OM. As
we have shown above, however, the Kubo-Streda formula and the semiclassical
formula give the same expression for the intrinsic OM. So no discrepancy is
expected to occur between the two approaches. The non-zero OM in Ref.
\cite{Shindou2001} in the Mott gap was ascribed by the authors to be a bulk
Fermi sea property. However, the present explicit relation Eq. (\ref{e23})
together with Haldane's argument on the metallic AHE \cite{Haldane2004} shows
that the non-quantized part of the OM in the present model is a Fermi-surface
Berry phase effect, while the quantized part is completely determined by the
topology of the filling bands.%

\begin{figure}[tbp]
\begin{center}
\includegraphics[width=0.7\linewidth]{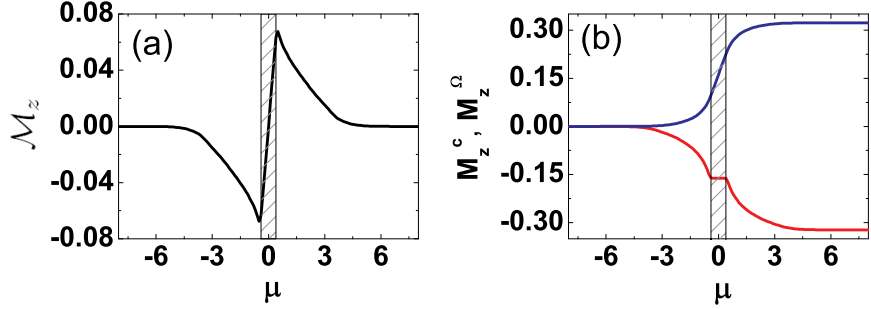}
\end{center}
\caption{(Color online). (a) Orbital magnetization $\mathcal{M}_{z}$, and (b)
its two components $M_{c}^{z}$ (red curve) and $M_{\Omega}^{z}$ (blue curve)
as a function of the electron chemical potential $\mu$ for distortion
$d$=$0.2$. The shaded area is the gap between the lower and upper
bands. To suppress the divergence at band/gap contact, we have used the
temperature of $k_{B}T$=$0.05$.}
\label{fig5}
\end{figure}%
Figure 5(a) shows the $\mathcal{M}_{z}$\ as a function of the electron
chemical potential $\mu$ for the distortion $d$=$0.2$. One can see that
initially the OM rapidly decreases as the filling of the lower bands
increases, arriving at a minimum at $\mu$=$-0.4$, a value corresponding to the
top of the lower band. Then, as the chemical potential continues to vary in
the gap [shaded region in Fig. 5(a)] between the lower and upper bands, the OM
goes up and increases as a linear function of $\mu$. This linear relationship
in the insulating region is explicit from Eq. (\ref{e22}). When the chemical
potential touches the bottom of the upper band, then the linear increase in
$\mathcal{M}_{z}$ suddenly stops and the OM rapidly decreases again by going
the chemical potential through the upper bands. The turning behavior at the
band/gap contacts becomes numerically divergent at $k_{B}T$=$0$. This
discontinuity is due to the singular behavior of $\Omega_{n}(\mathbf{k})$ at
the BZ edge point $\mathbf{k}$=$\mathbf{k}_{0}$, which will play its role when
the $k$-integral is over the entire BZ.

The distinct behavior of the OM in the metallic and insulating regions, as
shown in Fig. 5(a), reflects the different roles that its two components
$M_{c}^{z}$ and $M_{\Omega}^{z}$ play in these two regions. To see this, we
show in Fig. 5(b) $M_{c}^{z}$ (blue curve) and $M_{\Omega}^{z}$ (red curve) as
a function of $\mu$. One can see that $M_{c}^{z}$ and $M_{\Omega}^{z}$ oppose
each other, which implies that they are carried by opposite-circulating
currents. Also one can see that in the band insulating regime, the
conventional term $M_{c}^{z}$ keeps a constant during variation of $\mu$. This
behavior is due to the fact that the upper limit of the $k$-integral of
$\mathbf{m}_{n}(\mathbf{k})$ is invariant as the chemical potential varies in
the band gap. In the metallic region, however, since the occupied states
varies with the chemical potential $\mu$, thus $\mathbf{M}_{c}$ also varies
with $\mu$, resulting in a decreasing slope shown in Fig. 5(b). The Berry
phase term $\mathbf{M}_{\Omega}$ also displays different features in the
insulating and metallic regions. In the insulating region, $\mathbf{M}%
_{\Omega}$ linearly increases with $\mu$, as is expected from Eq. (\ref{e22}).
In the metallic region, however, this term keeps a constant with the amplitude
sensitively depending on the topological property of the band in which the
chemical potential is located. On the whole it reveals in Fig. 3 that the
metallic behavior of the OM is dominated by its conventional term
$\mathbf{M}_{c}$, while in the insulating regime the Berry phase term
$\mathbf{M}_{\Omega}$ comes to play a main role in determining the behavior of
the OM.

The above separate discussion of $\mathbf{M}_{c}$ and $\mathbf{M}_{\Omega}$
can be transferred to study the anomalous Nernst effect (ANE). The relation
between the OM and ANE has been recently found \cite{Xiao2}. To discuss the
transport measurement, it is important to discount the contribution from the
magnetization current, a point which has attracted much discussion in the
past. Cooper et al. \cite{Cooper} have argued that the magnetization current
cannot be measured by conventional transport experiments. Xiao et al.
\cite{Xiao2} have adopted this point and built up a remarkable picture that
the conventional orbital magnetic moment $\mathbf{M}_{c}$ does not contribute
to the transport current, while the Berry phase term in Eq. (\ref{e3})
directly enters and therefore modifies the intrinsic Hall current as follows%
\begin{equation}
\mathbf{j}_{\text{H}}\mathbf{=}-\frac{e^{2}}{\hbar}\mathbf{E}\times\sum
_{n}\int\frac{d^{3}k}{(2\pi)^{3}}f_{n}(\mathbf{r},\mathbf{k})\Omega
_{n}(\mathbf{k})\mathbf{-}\nabla\times\mathbf{M}_{\Omega}(\mathbf{r}),
\label{e24}%
\end{equation}
In the case of uniform temperature and chemical potential, obviously, the
second term is zero and the Hall effect of the distorted fcc lattice is
featured by the first term in Eq. (\ref{e24}). In the following, however, we
turn to study another situation, where the driving force is not provided by
the electric field. Instead, it is provided by a statistical force, i.e., the
gradient of temperature $T$. In this case, Eqs. (\ref{e24}) and (\ref{e3})
give the expression of intrinsic thermoelectric Hall current as $j_{x}%
=\alpha_{xy}(-\nabla_{y}T)$, where the anomalous Nernst conductivity
$\alpha_{xy}$ is given by%
\begin{align}
\alpha_{xy}  &  =\frac{1}{T}\frac{e}{\hbar}\sum_{n}\int\frac{d^{3}k}%
{(2\pi)^{3}}\Omega_{n}\label{e25}\\
&  \times\left[  \left(  \epsilon_{n\mathbf{k}}-\mu\right)  f_{n}+k_{B}%
T\ln\left(  1+e^{-\beta(\epsilon_{n\mathbf{k}}-\mu)}\right)  \right]
.\nonumber
\end{align}
%

\begin{figure}[tbp]
\begin{center}
\includegraphics[width=0.5\linewidth]{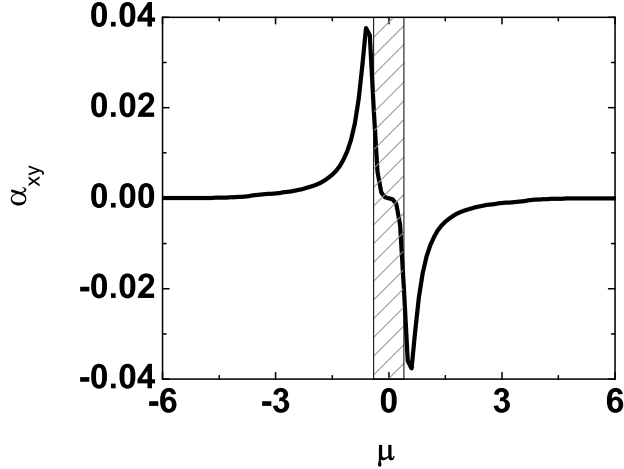}
\end{center}
\caption{The intrinsic anomalous Nernst conductivity $\alpha_{xy}%
$ as a function of the electron chemical potential $\mu$ for
$d$=$0.2$ and $k_{B}T$=$0.05$. The shaded area is the gap between the lower
and upper bands.}
\label{fig6}
\end{figure}%
Figure 6 shows $\alpha_{xy}$ of the distorted fcc lattice as a function of the
chemical potential for $d=0.2$ and $k_{B}T$=$0.05$. One can see that the ANE
disappears in the insulating regions, and when scanning $\mu$ through the
contacts between the bands and gaps, there will appear peaks and valleys.
Remarkably, a similar peak-valley structure was also found by the recent
first-principles calculations in CuCr$_{2}$Se$_{4}$ compound \cite{Xiao2}. The
ANE of this compound was recently measured by Lee et al. \cite{Lee} as a
function of Br doping which substitutes Se in the compound and changes the
chemical potential $\mu$. Due to the scarce data available, until now the
peak-valley structure of $\alpha_{xy}$ revealed in Fig. 6 and in Ref.
\cite{Xiao2} has not been found in experiment, and more direct experimental
results are needed for quantitative comparison with the theoretical results.
Interestingly, the expression for $\alpha_{xy}$ can be simplified at low
temperature as the Mott relation \cite{Xiao2},
\begin{equation}
\alpha_{xy}=-\frac{\pi^{2}}{3}\frac{k_{B}^{2}T}{e}\frac{\partial\sigma
_{xy}(\mu_{0})}{\partial\mu_{0}}. \label{e26}%
\end{equation}
Thus one can see that the low-temperature nonzero ANE is a Fermi-surface Berry
phase effect. Another unique feature of $\alpha_{xy}$ is its linear dependence
of temperature.

\section{Conclusion}

In summary, after pointing out the equivalence of the semiclassical approach
and the quantum Kubo-Streda formula in description of the intrinsic orbital
magnetization, we have theoretically studied the properties of the OM in
antiferromagnets on the distorted 3D fcc lattice. The distortion parameter
$d$\ in the fcc lattice produces nonzero 2D Chern number and results in
profound effects on the OM properties. An explicit relation between the OM and
the Hall conductivity in this system has been derived. According to this
relation we have found that the OM vanishes when the electron chemical
potential is lies in the Mott gap, which is in contrast with the results in
Ref. \cite{Shindou2001}. We have shown that the two parts $\mathbf{M}_{c}$ and
$\mathbf{M}_{\Omega}$ in the OM oppose each other, and yield the paramagnetic
and diamagnetic responses, respectively. In particular, due to its Fermi-sea
topological property, the magnetic susceptibility of $\mathbf{M}_{\Omega}$
remains to be a nonzero constant when the Fermi energy is located in the
energy gap. It has been further shown that the OM displays distinct behavior
in the metallic and Chern-insulating regions, because of different roles
$\mathbf{M}_{c}$ and $\mathbf{M}_{\Omega}$ play in these two regions. The
anomalous Nernst conductivity has been studied, which displays a peak-valley
structure as a function of the electron chemical potential. We expect that
these results will be experimentally verified in the spin-frustrated systems.

\begin{acknowledgments}
This work was supported by CNSF under Grant No. 10544004 and 10604010.
\end{acknowledgments}

\end{document}